# Fresnel diffraction in an interferometer: application to MATISSE

S. Robbe-Dubois*[a], Y. Bresson[b], E. Aristidi[a], S. Lagarde[b], P. Antonelli[b], B. Lopez[b], R. G. Petrov[a]

[a] Laboratoire Fizeau UMR 6525, Université de Nice Sophia-Antipolis, Observatoire de la Côte d'Azur, CNRS, Parc Valrose, 06108 Nice Cedex 2, France;

[b] Laboratoire Fizeau UMR 6525, Université de Nice Sophia-Antipolis, Observatoire de la Côte d'Azur, CNRS, Boulevard de l'Observatoire, B.P. 4229, 06304 Nice Cedex 4, France

## ABSTRACT

While doing optical study in an instrument similar to the interferometers dedicated to the Very Large Telescope (VLT), we have to take care of the pupil and focus conjugations. Modules with artificial sources are designed to simulate the stellar beams, in terms of collimation and pupil location. They constitute alignment and calibration tools. In this paper, we present such a module in which the pupil mask is not located in a collimated beam thus introducing Fresnel diffraction. We study the instrumental contrast taking into account the spatial coherence of the source, and the pupil diffraction. The considered example is MATISSE, but this study can apply to any other instrument concerned with Fresnel diffraction.

## 1. INTRODUCTION

The Multi AperTure mid-Infrared SpectroScopic Experiment[1] (MATISSE) is designed to be a spectro-interferometer combining the beams of up to four telescopes UTs/ATs of the European Southern Observatory VLT. Covering the L, M, and N bands, this second generation instrument has the scientific objectives to study Active Galactic Nuclei, the formation and evolution of planetary systems, the birth of massive stars as well as the observation of the high-contrast environment of evolved stars. The optics of MATISSE is divided in three main sub-systems: the Warm Optics (WOP) and two Cold Optics (COB) benches for the L&M and N bands[2]. To simulate the VLT beams, the MATISSE instrument can be lit by artificial sources (ARC). This module delivers four visible and IR collimated beams in the purpose of alignment, maintenance and calibration operations.

The purpose of this paper is to study the spatial coherence of the ARC. In order to optimize the space dedicated to this module, the optical configuration is such that the pupil mask is not in a collimated beam. In the frame of the instrument performance study[3], we were then concerned by the diffraction of the pupil mask, and its effect on the instrumental contrast measured in the Point Spread Function (PSF) plane. In Section 2 of this paper, we present the ARC optical configuration. Section 3 shows the Fresnel formalism[4] used to estimate the ARC performance in terms of instrumental fringe contrast. Section 4 is a study of a defocus effect, i.e. the contrast evolution if the observation is performed away from the theoretical PSF plane. The study is performed at the lowest wavelength of MATISSE, i.e. 3 μm.

## 2. ARC OPTICAL CONFIGURATION

The main function of the ARC module is to provide four beams at the entrance of the WOP to simulate the VLT beams to perform MATISSE alignment and calibration. The source is either a laser diode or an infrared ceramic source. Figure 1 shows the optical scheme of the ARC, from the pinhole source to the beam division.

The numerical values of the different parameters are the following:

- The power optics has a focal length $f_1$ = 300 mm. Its position defines the origin of the z-axis. We note $(x_1, y_1)$ the coordinates of a point in this $P_1$-plane.

- The extended source, a uniform, spatially incoherent disk of diameter $\rho$ = 25 μm, is located at a distance $z_1 = 2 f_1$ = 600 mm in front of the previous optics.

- We observe the figure of diffraction in an *x-y* (P) plane located at the distance $z_3 = z_1$ of the power optics. Without the PMA-mask it would be the image of the source.
- The PMA-mask is located in the $P_2$-plane at the distance $z_2 = 499$ mm of the power optics, and at the distance $z = 101$ mm of the P-plane containing the figure of diffraction. We note $(x_2, y_2)$ the coordinates of a point in this plane. The pupil diameter *d* is 3.1 mm, and their separation *b* is 5 mm.

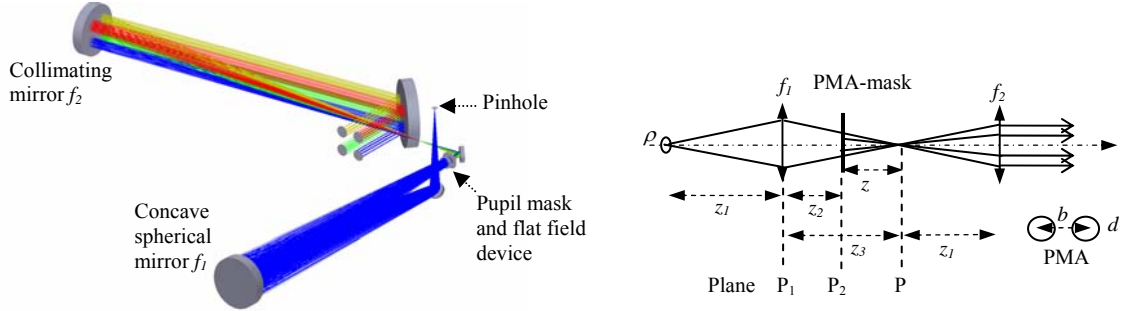

Figure 1. Left: optical scheme of the artificial source (ARC) module of MATISSE, from the pinhole to the 4 beam division. Right: simple view of the optical configuration of the ARC. The source, of diameter $\rho$, is located at a distance $z_1$ of a $P_1$-plane containing the power optics of focal length $f_1$. We observe the (Point Spread Function) PSF in a P-plane located at a distance $z_3$. The pupil mask PMA is at a distance *z* of the PSF. The pupil separation is *b*, and their diameter is *d*. A second optics, with focal length $f_2$, forms the collimated beams with a *B*-separation, and allows controlling the fringe pattern in the final instrument focal plane.

## 3. SPATIAL COHERENCE - CONTRAST EVALUATION

Let us consider an off-axis point-like source. Its position is described by the angular directions $\alpha$ and $\beta$ as seen from the origin. We note $\vec{\theta} = (\alpha, \beta)$. A diverging spherical wave is generated from this point. To describe the wave propagation along the *z*-direction, we use the Fresnel formalism[5] which is a convolution product * between the complex amplitude of the wave and a term noted $\frac{1}{i\lambda z} exp\left(\frac{i2\pi r}{\lambda}\right)$, where $\lambda$ is the wavelength and $r \approx z + \frac{(x^2 + y^2)}{2z}$ in the paraxial approximation. While the wave encounters optical elements, we multiply its amplitude by their transmission coefficients. The expression of the wave amplitude $\Psi$ in the P-plane is derived below considering arbitrary values for $z_1$, $f_1$ and $z_3$. The P-plane being the conjugate of the source plane, we use the expression valid in Gauss conditions $z_3 = \frac{z_1 f_1}{z_1 - f_1}$.

The wave amplitude $\Psi_{0-}(\vec{\rho}_1)$ incident on the $P_1$-plane, with $\vec{\rho}_1 = (x_1, y_1)$, is given by:

$$\Psi_{0-}(\vec{\rho}_1) = \frac{\psi_0}{z_1} \, exp\left(\frac{i2\pi}{\lambda} z_1\right) \, exp\left(\frac{i\pi}{\lambda z_1}(\vec{\rho}_1 - \vec{\theta} z_1)^2\right) \quad (1)$$

From the $P_1$-plane the wave goes through the optics with a transmission coefficient $exp\left(-\frac{i\pi \vec{\rho}_1^{\,2}}{\lambda f_1}\right)$ and propagates along a distance $z_2$. At the entrance of the PMA-mask the wave amplitude $\Psi_{z2-}(\vec{\rho}_2)$, with $\vec{\rho}_2 = (x_2, y_2)$, is:

$$\Psi_{z2-}(\vec{\rho}_2) = \left[\frac{\psi_0}{z_1} \, exp\left(\frac{i2\pi z_1}{\lambda}\right) exp\left(\frac{i\pi z_1 \vec{\theta}^{\,2}}{\lambda}\right) exp\left(-\frac{i\pi \vec{\rho}_1^{\,2}}{\lambda z_3}\right) exp\left(-\frac{i2\pi \vec{\theta}.\vec{\rho}_1}{\lambda}\right)\right] * \left[\frac{exp\left(\frac{i2\pi z_2}{\lambda}\right)}{i\lambda z_2} exp\left(\frac{i\pi \vec{\rho}_2^{\,2}}{\lambda z_2}\right)\right] \quad (2)$$

By developing the convolution product, we recognize the Fourier Transform (FT) of $exp\left(\frac{i\pi}{\lambda}\left(\frac{1}{z_2}-\frac{1}{z_3}\right)\vec{\rho_1}^2\right)$ with the conjugated variable $\frac{\vec{\rho_2}+\vec{\theta}z_2}{\lambda z_2}$. This results in:

$$\Psi_{z2\text{-}}(\vec{\rho_2}) = \frac{\psi_0 z_3}{(z_3-z_2)z_1} exp\left(\frac{i2\pi(z_1+z_2)}{\lambda}\right) exp\left(-\frac{i\pi \vec{\rho_2}^2}{\lambda(z_3-z_2)}\right) exp\left(\frac{i\pi(z_1 z_3 - z_1 z_2 - z_3 z_2)\vec{\theta}^2}{\lambda(z_3-z_2)}\right)$$

$$exp\left(-\frac{2i\pi z_3 \vec{\rho_2}.\vec{\theta}}{\lambda(z_3-z_2)}\right) \qquad (3)$$

$\Psi_{z2\text{-}}(\vec{\rho_2})$ represents a spherical wave in the P$_2$-plane having propagated a distance of $z_1+z_2$, and converging at $(-z_3\alpha, -z_3\beta, z_3-z_2)$ after the PMA.

The transmission coefficient $pup(\vec{\rho_2})$ of the PMA-mask is described by the following expression:

$$pup(\vec{\rho_2}) = \Pi\left(\frac{|\vec{\rho_2}|}{d}\right) * \left[\delta(\vec{\rho_2}-\vec{b}/2) + \delta(\vec{\rho_2}+\vec{b}/2)\right] \qquad (4)$$

$\Pi$ being the pupil function and $\delta$ the Dirac function. The parameters $\vec{b}$ and $d$ are respectively the vector between the two PMA pupils, and their diameter.

Taking into account the propagation through the distance $z = z_3 - z_2$, we find the expression of the wave amplitude $\Psi(\vec{\rho})$ in the P-plane, with $\vec{\rho} = (x, y)$:

$$\Psi(\vec{\rho}) = \frac{\psi_0 z_3}{i\lambda z_1(z_3-z_2)^2} exp\left(\frac{i2\pi(z_1+z_3)}{\lambda}\right) exp\left(\frac{i\pi \vec{\rho}^2}{\lambda(z_3-z_2)}\right) exp\left(\frac{i\pi(z_1 z_3 - z_1 z_2 - z_3 z_2)\vec{\theta}^2}{\lambda(z_3-z_2)}\right)$$

$$\iint exp\left(-\frac{2i\pi z_3 \vec{\rho_2}.\vec{\theta}}{\lambda(z_3-z_2)}\right) pup(\vec{\rho_2}) exp\left(-\frac{i2\pi}{\lambda(z_3-z_2)}\vec{\rho}.\vec{\rho_2}\right) d^2\vec{\rho_2} \qquad (5)$$

We recognize the FT of $exp\left(-\frac{2i\pi z_3 \vec{\rho_2}.\vec{\theta}}{\lambda(z_3-z_2)}\right) pup(\vec{\rho_2})$ with the conjugated variable $\frac{\vec{\rho}}{\lambda(z_3-z_2)}$. This results in:

$$\Psi(\vec{\rho}) = \frac{\psi_0 z_3}{i\lambda z_1(z_3-z_2)^2} exp\left(\frac{i2\pi(z_1+z_3)}{\lambda}\right) exp\left(\frac{i\pi \vec{\rho}^2}{\lambda(z_3-z_2)}\right) exp\left(\frac{i\pi(z_1 z_3 - z_1 z_2 - z_3 z_2)\vec{\theta}^2}{\lambda(z_3-z_2)}\right)$$

$$\delta\left(\frac{\vec{\rho}+\vec{\theta}z_3}{\lambda(z_3-z_2)}\right)\left[\frac{\pi d^2}{2} J_{1c}\left(\frac{\pi d}{\lambda(z_3-z_2)}|\vec{\rho}|\right) cos\left(\frac{\pi \vec{b}.\vec{\rho}}{\lambda(z_3-z_2)}\right)\right] \qquad (6)$$

Finally, the intensity of the diffraction figure in the P-plane is proportional to:

$$I(\vec{\rho}) \# \left|\delta(\vec{\rho}+\vec{\theta}z_3) * \left[J_{1c}\left(\frac{\pi d}{\lambda(z_3-z_2)}|\vec{\rho}|\right) cos\left(\frac{\pi \vec{b}.\vec{\rho}}{\lambda(z_3-z_2)}\right)\right]\right|^2$$

$$\# \left(J_{1c}\left(\frac{\pi d}{\lambda(z_3-z_2)}|\vec{\rho}-\vec{\theta}z_3|\right)\right)^2 \left(1+cos\left(\frac{2\pi \vec{b}.(\vec{\rho}-\vec{\theta}z_3)}{\lambda(z_3-z_2)}\right)\right) \qquad (7)$$

where $J_1$ is the Bessel function of the first kind and $J_{1c}(q)=J_1(q)/q$.

Let us remember that $I(\vec{\rho})$ is the intensity formed by a single point source. It describes a PSF modulated by interference fringes with a contrast equal to 1. The Airy disk size is $2.44\lambda(z_3-z_2)/d$. The $\alpha z_3$ and $\beta z_3$ terms represent the PSF and fringe shift depending on the point source position. Let us now consider the extended source of surface $S$ as a set of incoherent points, each one emitting a spherical wave. The resulting PSF in the P-plane is then the sum of the intensities of each wave. In the hypothesis the effect of the displacement $\alpha z_3$ is negligible, it can be approximated by:

$$PSF(\vec{\rho}) \# \iint_S I(\vec{\rho}) d^2\vec{\theta} \# \left(J_{1c}\left(\frac{\pi d}{\lambda z}|\vec{\rho}|\right)\right)^2 \left\{1 + 2 J_{1c}\left(\frac{\pi \rho b}{\lambda z}\right) \cos\left(\frac{2\pi \vec{b}\cdot\vec{\rho}}{\lambda z}\right)\right\} \quad (8)$$

with $\rho$ the source diameter, and $z = z_3 - z_2$. This approximation is valid if we consider the maximum value of $\alpha z_3 = z_3 \rho/z_1 \# 25$ µm, and the $J_{1c}^2$ size of $2.44\lambda z/d \# 250$ µm. The fringe contrast is then simply $2 J_{1c}\left(\frac{\pi \rho b}{\lambda z}\right)$ which is the FT of the pupil function. The result is identical to that obtained when the pupil mask is illuminated with a collimated beam in a configuration similar to that shown in Figure 2. The mask must present a size homothetic to the initial one such that the pupil diameter $d'$ is $dz_3/z$, and the separation $b'$ is $bz_3/z$.

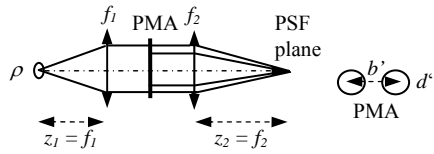

Figure 2. Equivalent optical configuration of the ARC. The PMA, of dimensions $d'$ and $b'$, is located in a collimated beam.

Considering the parameter values of MATISSE (see Section 2), we can plot the contrast versus $\rho$ for a given $z$, and the contrast versus $z$ for a given $\rho$ (Figure 3). With $\rho = 25$ µm and $z = 101$ mm, the expected instrumental contrast, calculated without the approximation of eq. 8, is 0.76 at the wavelength of 3 µm. Around $z = 101$ mm, an error $\Delta z$ of ±2 mm on the PMA-mask positioning introduces less than 1% change in the contrast value.

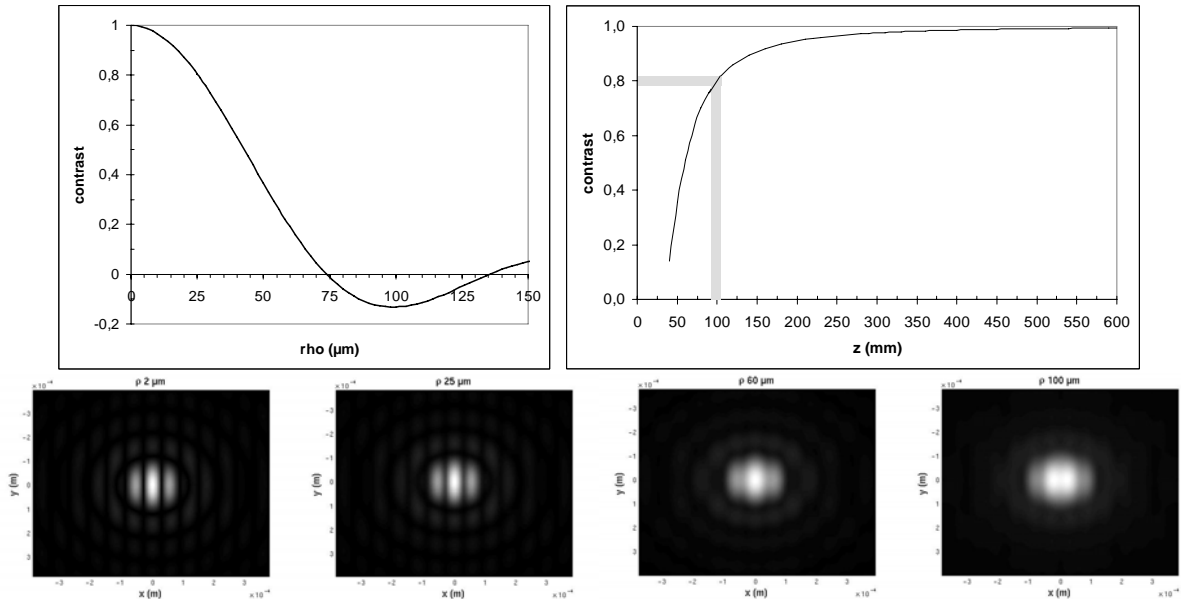

Figure 3. Top left: instrumental contrast versus the source diameter $\rho$ with $z = 101$ mm (MATISSE configuration) at $\lambda = 3$ µm. The PSF shift is here neglected. Top right: instrumental contrast versus the distance $z$ between the mask and the PSF plane with $\rho = 25$ µm (MATISSE source) at $\lambda = 3$ µm. Bottom: simulated PSFs for four values of $\rho$ with $z = 101$ mm. The contrast obtained from the simulations is 1 for $\rho = 2$ µm, 0.76 for $\rho = 25$ µm, 0.17 for $\rho = 60$ µm, and 0.15 with a contrast inversion for $\rho = 100$ µm.

# 4. DEFOCUS EFFECT

We study the effect if we observe in a plane located at a distance $\varepsilon$ from the theoretical PSF P-plane. To simplify, let us consider $z_1 = z_3$ as in the MATISSE instrument. In this case, the wave propagates on a distance $z = z_1 - z_2 - \varepsilon$ from the PMA-mask up to the plane of observation. The amplitude $\Psi(\vec{\rho})$ is then the amplitude at the exit of the PMA convolved by $\exp\left(\frac{i2\pi r}{\lambda}\right) / (i\lambda z)$, with $r \# z_1-z_2 + \rho^2/2z$.

$$\Psi(\vec{\rho}) = \frac{\psi_0}{i\lambda z(z_1-z_2)} \exp\left(\frac{i2\pi}{\lambda}(2z_1-\varepsilon)\right) \exp\left(\frac{i\pi}{\lambda}\left(\frac{z_1(z_1-2z_2)}{z_1-z_2}\right)\vec{\theta}^2\right) \exp\left(\frac{i\pi\vec{\rho}^2}{\lambda z}\right)$$

$$\iint \exp\left(\frac{-2i\pi z_1 \vec{\rho}_2.\vec{\theta}}{\lambda(z_1-z_2)}\right) pup(\vec{\rho}_2) \exp\left(\frac{i\pi\varepsilon\vec{\rho}_2^2}{\lambda(z_1-z_2)z}\right) \exp\left(\frac{-i2\pi}{\lambda z}\vec{\rho}.\vec{\rho}_2\right) d^2\vec{\rho}_2 \tag{9}$$

We recognize the FT of $\exp\left(\frac{-2i\pi z_1 \vec{\rho}_2.\vec{\theta}}{\lambda(z_1-z_2)}\right) pup(\vec{\rho}_2) \exp\left(\frac{i\pi\varepsilon\vec{\rho}_2^2}{\lambda(z_1-z_2)z}\right)$ with the conjugated variable $\frac{\vec{\rho}}{\lambda z}$.

This is also the convolution product of $\delta\left(\frac{\vec{\rho}+\vec{\theta}z_1}{\lambda(z_1-z_2)}\right) * \left[\frac{\pi d^2}{2} J_{1c}\left(\frac{\pi d}{\lambda(z_1-z_2)}|\vec{\rho}|\right) \cos\left(\frac{\pi \vec{b}.\vec{\rho}}{\lambda(z_1-z_2)}\right)\right]$

with $\frac{i\lambda(z_1-z_2)z}{\varepsilon} \exp\left(\frac{-i\pi(z_1-z_2)\vec{\rho}^2}{\lambda \varepsilon z}\right)$.

The expression of $\Psi(\vec{\rho})$ can be simplified for small value of $\varepsilon$.

To lighten the expression in the following, let's call $\psi_0'$ the term multiplying the integral. The Taylor series at first order of $\Psi(\vec{\rho})$ results in:

$$\Psi(\vec{\rho}) \# \psi_0' \iint \exp\left(-\frac{2i\pi z_1 \vec{\rho}_2.\vec{\theta}}{\lambda(z_1-z_2)}\right) pup(\vec{\rho}_2) \exp\left(\frac{-i2\pi}{\lambda z}\vec{\rho}.\vec{\rho}_2\right) d^2\vec{\rho}_2$$

$$+ \frac{i\pi\psi_0'\varepsilon}{\lambda(z_1-z_2)z} \iint \exp\left(\frac{-2i\pi z_1 \vec{\rho}_2.\vec{\theta}}{\lambda(z_1-z_2)}\right) pup(\vec{\rho}_2)\vec{\rho}_2^2 \exp\left(\frac{i\pi\varepsilon\vec{\rho}_2^2}{\lambda(z_1-z_2)z}\right) \exp\left(\frac{-i2\pi}{\lambda z}\vec{\rho}.\vec{\rho}_2\right) d^2\vec{\rho}_2 \tag{10}$$

For small values of $\varepsilon$, the first term is equivalent to the amplitude for $\varepsilon = 0$. Let us call it $\Psi_{\varepsilon=0}(\vec{\rho})$.

The second term is proportional to the FT of $\vec{\rho}_2^2 f(\vec{\rho}_2)$, with conjugated variable $\frac{\vec{\rho}}{\lambda z}$, and with

$$f(\vec{\rho}_2) = \exp\left(\frac{-2i\pi z_1 \vec{\rho}_2.\vec{\theta}}{\lambda(z_1-z_2)}\right) pup(\vec{\rho}_2).$$

We know that $FT_{\vec{w}}(\vec{\rho}_2^2 f(\vec{\rho}_2)) = -\frac{1}{4\pi^2}\left(\frac{\delta^2 \hat{f}(\vec{w})}{\delta u^2} + \frac{\delta^2 \hat{f}(\vec{w})}{\delta v^2}\right)$, with $\hat{f}(\vec{w})$ the FT of the function $f$ with conjugated variable $\vec{w} = (u, v)$, and $\hat{f}(\vec{w}) = \delta\left(\vec{w} + \frac{\vec{\theta}z_1}{\lambda z}\right) * \left[\frac{\pi d^2}{2} J_{1c}(\pi d |\vec{w}|).\cos(\pi\vec{b}.\vec{w})\right]$.

We are interested in the contrast loss due to $\varepsilon$. The intensity modulation being defined in the x-direction, let us consider $v = 0$ ($y = 0$). To simplify also the expression, we consider $\beta = 0$ such that the position of the point-like source is defined by the angle $\alpha$. We checked the analytical results below with computer algebra, using the MAPLE software from Maplesoft society, Canada.

To calculate the term $FT_{u,v}(x_2^2.f(x_2,y_2))$, we used the mathematical derivatives concerning the Bessel functions of the $n^{th}$ kind, and considering a constant $m$ in the argument:

$$\frac{d[v^{-n}J_n(mv)]}{dv} = -m.v^{-n}J_{n+1}(mv) \text{ and } \frac{d[J_n(mv)]}{dv} = mJ_{n-1}(mv) - \frac{n}{v}J_n(mv)$$

$$\hat{f}(u,0) = \frac{d}{2}\delta\left(u+\frac{\alpha z_1}{\lambda z}\right) * \left[\frac{J_1(\pi du)}{u}\cos(\pi bu)\right] \text{ and}$$

$$\frac{d^2\hat{f}(u,0)}{du^2} = \frac{d}{2}\frac{d^2}{du^2}\left[\delta\left(u+\frac{\alpha z_1}{\lambda z}\right)*\left(\frac{J_1(\pi du)}{u}\cos(\pi bu)\right)\right] = \frac{d}{2}\frac{d^2}{dv^2}\left[\frac{J_1(\pi dv)}{v}\cos(\pi bv)\right] \text{ with } v=u-\frac{\alpha z_1}{\lambda z}.$$

$$\frac{d}{dv}\left[v^{-1}J_1(\pi dv)\cos(\pi bv)\right] = -\pi d\frac{J_2(\pi dv)}{v}\cos(\pi bv) - \pi b\frac{J_1(\pi dv)}{v}\sin(\pi bv)$$

$$\frac{d^2}{dv^2}\left[v^{-1}J_1(\pi dv)\cos(\pi bv)\right] = 3\pi d\frac{J_2(\pi dv)}{v^2}\cos(\pi bv) + 2\pi^2 db\frac{J_2(\pi dv)}{v}\sin(\pi bv) - \pi^2(d^2+b^2)\frac{J_1(\pi dv)}{v}\cos(\pi bv)$$

Considering the term $FT_{u,v}(y_2^2.f(x_2,y_2))$, we also used the following derivative:

$$\frac{d\left[J_{1c}\left(2\pi d\sqrt{u^2+v^2}\right)\right]}{dv} = -2\frac{J_1\left(2\pi d\sqrt{u^2+v^2}\right)v}{\pi d(u^2+v^2)^{3/2}} + \frac{J_0\left(2\pi d\sqrt{u^2+v^2}\right)v}{(u^2+v^2)}$$

The resulting amplitude $\Psi(u,0)$ is then:

$$\Psi(u,0) = \Psi_{\varepsilon=0}(u,0) - \psi_0'\frac{i\varepsilon d}{8\lambda(z_1-z_2)z}\delta\left(u+\frac{\alpha z_1}{\lambda z}\right) *$$

$$\left(3d\frac{J_2(\pi du)}{u^2}\cos(\pi bu) + 2\pi db\frac{J_2(\pi du)}{u}\sin(\pi bu) - \left(\pi(d^2+b^2)+\frac{2}{\pi u^2}\right)\frac{J_1(\pi du)}{u}\cos(\pi bu) + d\frac{J_0(\pi du)}{u^2}\cos(\pi bu)\right)$$

The resulting intensity is given by:

$$I(x,0) = \frac{|\psi_0|^2}{4\lambda^2 z^4}\delta(x+\alpha z_1) * \left\{\left(\pi d^2 J_{1c}\left(\frac{\pi dx}{\lambda z}\right)\right)^2\left(1+\cos\left(\frac{2\pi bx}{\lambda z}\right)\right)\right.$$

$$+\left(\frac{\varepsilon d}{4\lambda(z_1-z_2)z}\right)^2\left[3d\left(\frac{\lambda z}{x}\right)^2 J_2\left(\frac{\pi dx}{\lambda z}\right)\cos\left(\frac{\pi bx}{\lambda z}\right) + 2\pi db\frac{\lambda z}{x}J_2\left(\frac{\pi dx}{\lambda z}\right)\sin\left(\frac{\pi bx}{\lambda z}\right)\right.$$

$$\left.\left.-\left(\pi(d^2+b^2)+\frac{2}{\pi}\left(\frac{\lambda z}{x}\right)^2\right)\frac{\lambda z}{x}J_1\left(\frac{\pi dx}{\lambda z}\right)\cos\left(\frac{\pi bx}{\lambda z}\right) + d\left(\frac{\lambda z}{x}\right)^2 J_0\left(\frac{\pi dx}{\lambda z}\right)\cos\left(\frac{\pi bx}{\lambda z}\right)\right]^2\right\} \quad (11)$$

with $z = z_1 - z_2 - \varepsilon$.

This expression is valid for $\frac{\pi(x_2^2+y_2^2)\varepsilon}{\lambda(z_1-z_2)z} \ll 1$ (eq. 9), i.e. for $\varepsilon \ll 1$ mm considering that $(x_2^2+y_2^2) \leq d^2 \sim 2.4\mu m^2$.

Figure 4 shows the effect of $\varepsilon$ on $I(x,0)$ of eq. 11: the minima are not zero any more, the interfringe is changed. To enhance the illustration, we plotted the intensity for $\varepsilon = 1$ mm. The resulting contrast degradation factor is better than 0.96 for values of $\varepsilon$ up to 0.5 mm.

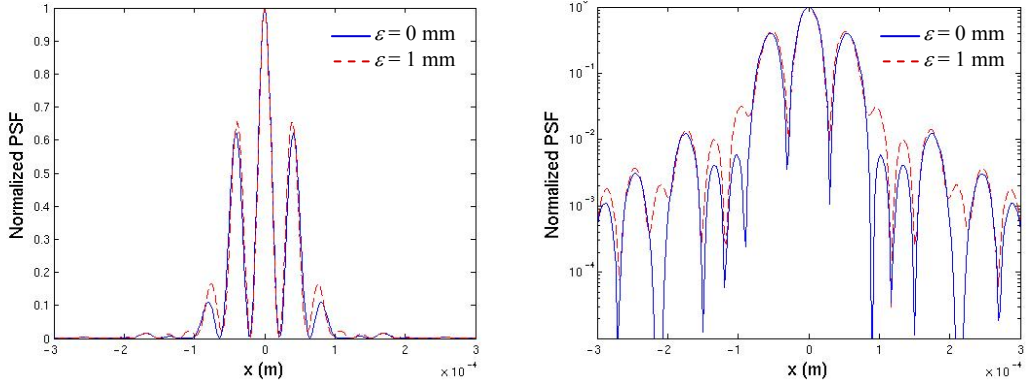

Figure 4. Normalized Intensity *I(x, 0)*, eq. 11, for $\varepsilon = 0$ mm (blue full line) and for $\varepsilon = 1$ mm (red dashed line). On the right figure, data are also plotted as logarithmic scale for the y-axis.

To have access to the contrast value versus $\varepsilon$ without the approximations considered to find eq. 11, we performed the direct computation of eq. 9. We validated the results by comparison with those of Section 3 for which $\varepsilon = 0$, and finally found the contrast degradation factor versus $\varepsilon$ shown on Figure 5. The contrast degradation factor is 0.95 for $\varepsilon = 1$ mm, and 0.40 for $\varepsilon = 3$ mm. The associated simulated PSFs are also shown.

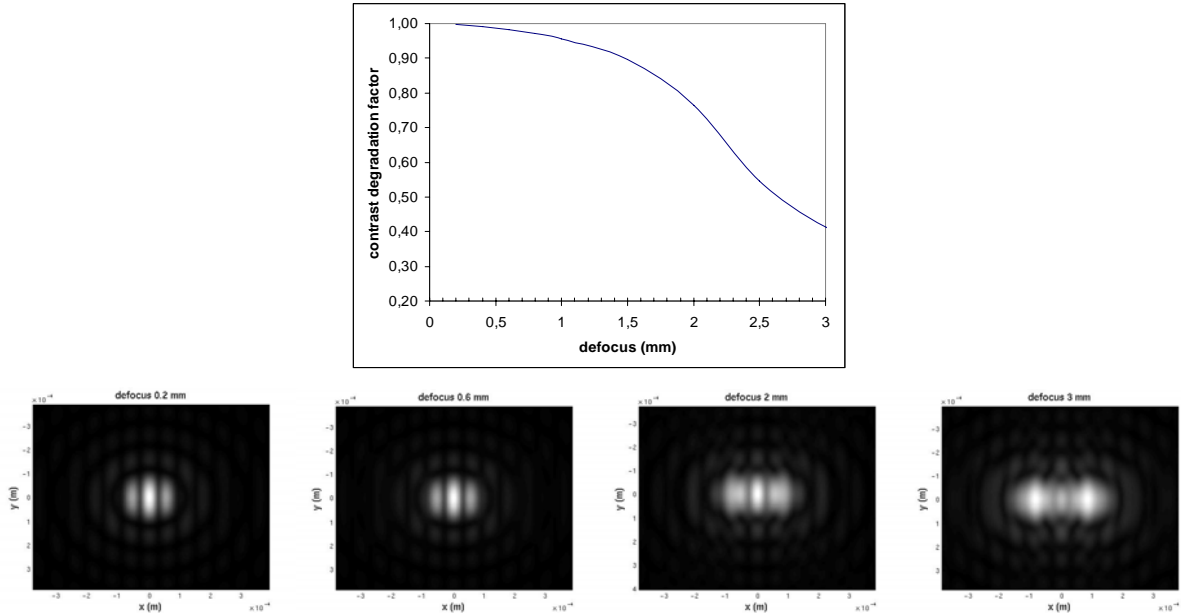

Figure 5. Top: degradation factor of the instrumental contrast versus the error $\varepsilon$ in the observation plane of the PSF. The computation is performed directly from eq. 9. The source diameter $\rho$ is 25 µm and $\lambda = 3$ µm. Bottom: simulated PSFs for four values of $\varepsilon$. The contrast degradation factor is 1 for $\varepsilon = 0.2$ mm, 0.98 for $\varepsilon = 0.6$ mm, 0.95 for $\varepsilon = 1$ mm, and 0.40 for $\varepsilon = 3$ mm.

## 5. CONCLUSION

We studied the spatial coherence of the artificial source module of MATISSE considering the pupil mask is not in a collimated beam. Using Fresnel formalism, we showed the PSF is the same than that obtained in an optical configuration with a pupil mask in a collimated beam. Considering the MATISSE parameters, the instrumental contrast delivered by this module is expected to be 0.76 at the wavelength of 3 µm. We then estimated that the contrast is degraded by a factor of 95% when the PSF is recorded in a plane located at a distance of 1 mm from the theoretical PSF plane. This allows estimating the specification of defocus for the final MATISSE detector.